\documentclass[doublecol]{epl2}

\usepackage{graphicx}
\usepackage{amsmath,amssymb,amsfonts,color}

\title{Direct observation of the tube model in F-actin solutions: \\
    Tube dimensions and curvatures}
\shorttitle{Direct observation of the tube model in F-actin solutions }

\author{M. Romanowska\inst{1,2} \and H. Hinsch\inst{3} \and N. Kirchge\ss ner\inst{1} \and M. Giesen\inst{1} \and M. Degawa\inst{4} \and B.
Hoffmann\inst{1} \and E. Frey\inst{3} \and R. Merkel\inst{1}} \shortauthor{M.
Romanowska \etal}

\institute{
  \inst{1} Institute of Bio- and Nanosystems
    4: Biomechanics, Research Centre J\"{u}lich, 52425 J\"{u}lich,
    Germany\\
  \inst{2} Marian Smoluchowski Institut of
    Physics, Jagiellonian University Krak\'ow, Reymonta 4, 30-059
    Krak\'ow, Poland\\
  \inst{3} Arnold Sommerfeld Center for
    Theoretical Physics and Center for Nanoscience,
    Ludwig-Maximilians-Universit\"{a}t M\"{u}nchen, 80333 Munich,
    Germany\\
  \inst{4} Laboratory of Molecular Dynamics, Brain Science Institute, RIKEN, Wako,
  Saitama 351-0198, Japan\\
} 
{Biopolymers} 

\abstract{ Mutual uncrossability of polymers generates topological
constraints on their conformations and dynamics, which are generally
described using the tube model. We imaged confinement tubes for
individual polymers within a F-actin solution by sampling over many
successive micrographs of fluorescently labeled probe filaments. The
resulting average tube width shows the predicted scaling behavior.
Unexpectedly, we found an exponential distribution of tube curvatures
which is attributed to transient entropic trapping in network void
spaces.}

\begin{document}

\maketitle

\section{Introduction}
One of the central questions of polymer physics is how the properties
  of polymeric materials, such as fluids or gels, derive from the
  characteristics of the individual macromolecules forming the material.
  In polymer solutions, entanglements between neighboring filaments
  impose topological constraints on the Brownian motion of individual
  polymers.  These constraints, however, are notoriously difficult to
  incorporate into a theoretical description  of
  the system.
  Arguably the most successful phenomenological approach to
  this problem is the \emph{tube model} introduced by
  Edwards~\cite{Edwards:1967}. It reduces the complex many-filament
  problem to a tractable single-filament problem, where the topological
  constraints imposed by the surrounding polymers are condensed to the
  concept of an impenetrable tube.  This assumption is at the heart of
  most theories of macromolecular behavior such as the reptation
  model~\cite{degennesscal}. For solutions of linear polymers that are
\emph{not}
  cross-linked the confinement tube is a dynamic entity that is
defined on
  intermediate time scales ranging from the moment of first contact
between
  neighboring polymers to the characteristic time for tube
remodelling. However,
  this time window can be of substantial size \cite{degennesscal}.

  Up to now the tube model has remained mostly a conceptual device for
  calculations with few direct experimental observations
  \cite{kassackmann1994,schleger:1998}.  Due to this sparseness of data
  most theories on collective properties of semi-dilute or dense polymer
  solutions rely on the assumption that filaments are enclosed in
  confinement tubes with fixed widths throughout the
  sample. Moreover, the ensemble of confinement tubes is usually assumed
  to qualitatively show the same conformation statistics as an
  individual polymer in free solution. In the present study, we
  challenged these assumptions by directly imaging confinement tubes of
  fluorescently labeled actin filaments embedded in semi-dilute
  solutions of unlabeled filaments.

  Actin filaments (F-actin) form by self-aggregation of the protein
  actin and exhibit lengths of some tens of microns. In solution, these
  filaments undergo thermal shape fluctuations but, due to their
  bending stiffness, maintain an average direction and are therefore
  referred to as semiflexible; their persistence length ranges from
  15\,--\,20~$\mu$m for phalloidin stabilized filaments
  \cite{ott:1993,gittes:1993,freytracer} to approximately 10~$\mu$m for
  unstabilized filaments \cite{isambertcarlier1995}. At present, the
  properties
  of isolated filaments are reasonably well understood in terms of the
  wormlike chain model \cite{RDFwilhelmfrey1996} but the relation
  between
  microstructure and viscoelastic properties of entangled solutions is
  still a challenging problem \cite{shinweitz2004,liu:2006}.

  For semiflexible polymers, a typical tube is pictured as a thin and
  relatively straight cylindrical pore. A thermally fluctuating filament
  confined within such a tube will encounter collisions with the tube
  wall, whose average distance is given by Odijk's deflection length
  \cite{odijk1983}. Expanding this concept Semenov derived a scaling law
  for the dependence of tube width, $w$, on monomer concentration, $c$
  \cite{semenov1986}. Basically, he exploited the fact that fluctuating
  filaments explore non-overlapping regions of space and predicted
  \begin{equation} \label{eq:scaling}
  w \propto c^{-3/5} \, .
  \end{equation}
  In contrast, the mean distance between neighboring filaments, $\xi$,
     i.e.\ the mesh size, scales for purely geometrical reasons like
  $c^{-1/2}$. Thus mesh size and confinement tube width are genuinely
  different length scales. Confinement is usually described
  by a harmonic tube potential which has been confirmed by experiments
  \cite{dichtl1999} and computer simulations \cite{hinsch07}.

\begin{figure}
           \onefigure{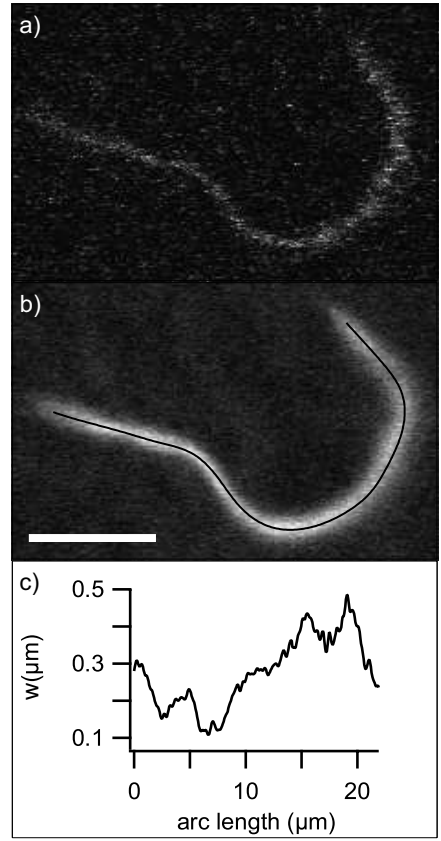}
                    \caption{(a) Micrograph of a fluorescently labeled F-actin
                    within a solution of concentration $c=$14.5~$\mu$M. (b) Average
                    of 100 successive images taken within 126 s. Overlayed is the
                    center line of the averaged contour (black line). Scale bar
                    5~$\mu$m. (c) The width of the averaged intensity distribution
                    corrected for optical resolution effects. Arc length runs from
                    left to right in the
                    micrographs.}
             \label{micrograph}
\end{figure}

In this Letter, we present a quantitative study of confinement tubes
  in entangled F-actin networks. We fluorescently labeled a trace amount
  of actin filaments with rhodamine-phalloidin, embedded those in a
  solution of unlabeled filaments and observed individual filaments by
  confocal laser scanning microscopy. A typical micrograph
  is shown in Fig.~\ref{micrograph}a. Upon averaging
  time series of such images, we obtained an intensity distribution
  representing the probability density of a filament within its tube
  (Fig. \ref{micrograph}b). This approach allowed us to directly
  characterize the size and conformation of individual confinement tubes
  in real space at optical resolution.

\section{Experimental}
Actin was prepared as described in the literature
  \cite{fletcherpollard1980}. Monomeric actin was kept in G-buffer (2.0
  mM Tris-HCl, pH 8.0, 0.2 mM CaCl$_{2}$, 0.2 mM ATP, 0.2 mM DTT) at
  4$^{\circ}$C and used for experiments within 72~h. Polymerization was
  initiated by adding 1/10 of the sample volume of 10$\times$ F-buffer
  (20 mM Tris-HCl, pH 7.5, 2 mM CaCl$_{2}$, 5 mM ATP, 2 mM DTT, 20 mM
  MgCl$_{2}$, 1 M KCl). The samples were then incubated for 2~h at
  37$^{\circ}$C. A part of the F-actin was labeled with
  rhodamine-phalloidin (TRITC-phalloidin) at a molar ratio of 1:1.
  Labeled and unlabeled F-actin solutions of the same concentration
  were then gently mixed at 1:1000 ratio for 15 min in a slow (few rpm)
  rotary shaker. The resulting solution was gently pipetted onto a micro
  slide and overlayed with a cover slide which was held at a distance of
  100 $\mu$m by a spacer of adhesive tape. Experiments were performed
  after an equilibration period of 15 min.

  Filaments were observed using an inverse confocal microscope (LSM510
  equipped with a water immersion objective, C-Apochromat 40/1.2W corr,
  Carl Zeiss, Jena, Germany). Excitation wavelength was 543 nm. Only
  filaments residing in a plane perpendicular to the optical axis at a
  distance of 40-60~$\mu$m from the cover slide were considered. The
  thickness of the optical slice was typically set to 2.8~$\mu$m. A
  field of view containing one labeled filament was scanned within
  0.1-0.3~s. Images were taken at 1~s intervals within a time period of
  2-3 minutes. On this time scale we observed only transverse filament
  fluctuations. Time-lapse sequences of about 25 filaments were
  collected for each concentration.

  As contour characterizing a confinement tube we used the center
  line of the intensity distribution in the time averaged micrographs,
  cf.  Fig.~\ref{micrograph}b. We determined this contour by the
  following algorithm: The image was binarized with a global threshold.
  The resulting areas were skeletonized to obtain a first estimate of
  the contour at pixel accuracy. Then the end points of this line were
  kept fixed and an active contour algorithm based on
  energy-minimization was applied on the time averaged micrographs
  as described in Ref.~\cite{amini1990}.
  This resulted in a smooth line representing the contour of the
  confinement tube; see Fig.~\ref{micrograph}b.  All image processing
  routines were developed using MatLab (Release 14, The Mathworks,
  Natick, MA). The width of the confinement tube, $\sigma_{\rm exp}$,
  was determined as the standard deviation of a Gaussian fit to the
  intensity distribution of the averaged image along lines perpendicular
  to the contour. A result from this procedure is shown in
  Fig.~\ref{micrograph}c.

\section{Experimental Results}
\begin{figure}
    \onefigure{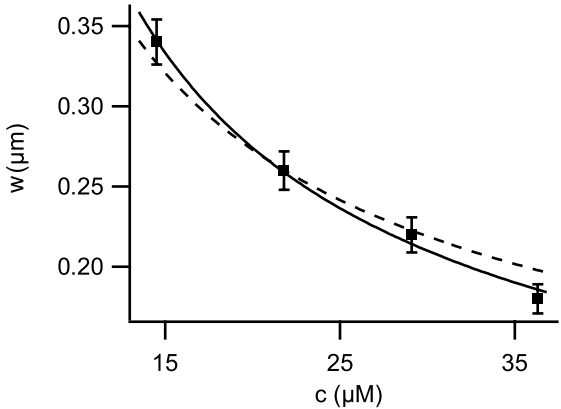}
    \caption{Dependence of the average tube widths, $w$, on actin
    concentration, $c$. Symbols: measured points, error bars: standard deviations of
    the mean tube widths of the filaments, full line: best fit to a power law
    (resulting exponent -0.61), dashed line: best fit with exponent -1/2.}
    \label{tubescaling}
  \end{figure}

Our first objective was to analyze the functional dependence of the
  average \emph{tube width} $w$ on the actin monomer concentration
  $c$. The data for the width $\sigma_{\rm exp}$ were corrected for the
  optical resolution\footnote{The optical resolution
    was calculated by convoluting the point spread function measured
on 100~nm fluorescent beads with a line.
    The resulting line spread function (LSF) was well approximated by a
    Gaussian of width $\sigma_{\rm LSF}$, which on average was
    0.18~$\mu$m}
     $\sigma_{\rm LSF}$  to give an undistorted tube width $w=(\sigma_{\rm
    exp}^2-\sigma_{\rm LSF}^2)^{1/2}$. We also took into account that a
  so-called critical concentration (1.9~$\mu$M) of actin monomers forms
  a pool of unpolymerized molecules in equilibrium with the actin
  filaments, i.e.\ in Eq.~\ref{eq:scaling} the concentration of actin
  within filaments was taken to be 1.9~$\mu$M less than the total actin
  concentration.  Average widths of the confinement tubes were 0.34,
  0.26, 0.22, and 0.18 $\mu$m at total actin concentrations of 14.5,
  21.8, 29.1, and 36.3~$\mu$M, respectively. These data conform quite
  well to a power law curve with a best fit for an exponent of -0.61,
  cf.\ Fig.~\ref{tubescaling}. An exponent -1/2, as expected for the
  mesh size, is clearly not compatible with our data. To our knowledge,
  this result is the first experimental confirmation of the predicted
  Semenov scaling exponent of -3/5, Eq.~\ref{eq:scaling}, showing that
  indeed tube width and mesh size are distinct length
  scales~\cite{semenov1986}.  Moreover, even the absolute values of the
  tube width as predicted by recent theoretical studies
  \cite{morse2001,hinsch07} are of comparable but lower magnitude. This
  is consistent since these treatments do not include collective
  fluctuations of the entangled solution and therefore give lower
  bounds.

So far our results nicely confirmed previous theoretical predictions on the average
tube width. However, contrary to common belief, fluctuations of the tube widths were
pronounced with standard deviations of $\approx$35\%; cf.\ Fig.~\ref{micrograph}c.
The length scale over which tube width fluctuations occurred was studied
via the autocorrelation function of $\sigma_{exp}$, defined as $ACF(j) = \frac{
\sum_i^N (\sigma_i - <\sigma>)(\sigma_{i+j} - <\sigma>) } {\sum_i^N (\sigma_i -
<\sigma>)(\sigma_{i} - <\sigma>)}$. In this equation the subscript ``exp'' was
dropped for the sake of brevity. The initial decay parameters
of the autocorrelation functions of $\sigma_{\rm exp}$ were 0.33, 0.49, 0.60,
0.68~$\mu$m$^{-1}$ at concentrations of  14.5, 21.8, 29.1, and 36.3~$ \mu$M,
respectively. Thus tube width fluctuations occurred on length scales of few microns.
The faster decay of correlations in the tube width at higher concentrations is a
consequence of the higher obstacle density.

\begin{figure}
    \onefigure{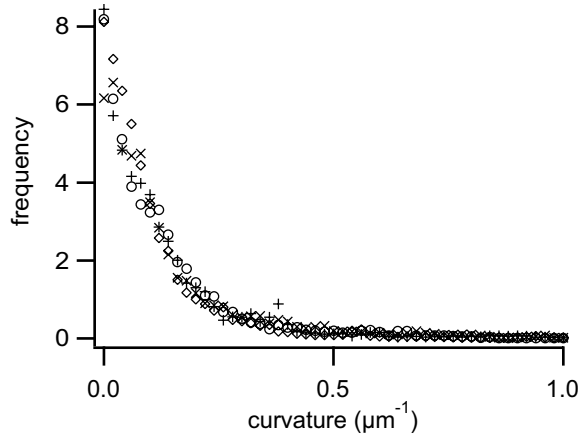}
      \caption{Distribution functions of the curvatures of the
        confinement tubes.  $+$ 36.3~$\mu$M, $\times$ 29.1~$\mu$M,
        $\circ $ 21.8~$\mu$M, and $ \diamond $ 14.5~$\mu$M.}
    \label{curvatures}
  \end{figure}

To go beyond a characterization of a tube in terms of its typical
  size we investigated tube conformations. Specifically, we asked for
  the probability distribution function of \emph{tube curvatures}.  In
  the absence of excluded volume it seems obvious that the probability
  distribution of polymer conformations in solution is identical to
  that of free filaments.  Sometimes, it is argued that even the
  (coarse-grained) conformations of the tube contour should follow
  such a free distribution \cite{doi1985}.  Since the effective free
  energy of the wormlike chain model is quadratic in the local
  curvature this would imply a Gaussian distribution for the
  curvatures of confinement tubes. To our great surprise we found a
  curvature distribution with a pronounced exponential tail; see
  Figs.~\ref{curvatures} and \ref{curv_theory}~\textit{top}.

The initial decay length of the curvature distribution shows little to no
dependence on concentration; the decay lengths of the distributions were 8.0, 8.0,
8.2, and
  9.9~$\mu$m at the concentrations of 36.3, 29.1, 21.8, and 14.5~$\mu$M,
  respectively. This is to be expected since small curvatures
  are dominated by bending stiffness.

  An exponential distribution decays much slower towards large values than a
Gaussian. This is reflected in our frequent observation of highly bent filaments.
For isolated filaments, similarly bent contours are extremely rare and were never
observed~\cite{ott:1993,gittes:1993,freytracer}. What is causing such a qualitative
change in chain conformations?  At this point several reasons for the occurrence of
highly bent shapes, besides topological interactions,  were conceivable. We could
rule out that filaments assumed such conformations during sample preparation and
remained kinetically trapped by additional experiments where the time delay between
successive  micrographs of actin filaments was set in the range from 10 to 30 sec.
With these longer time delays we were able to track individual filaments  for up to
one hour. On these time scales we observed several initially  straight filaments
going into highly bent shapes. This observation indicates that the highly bent
shapes causing the exponential decay of the distribution of curvatures of the
confinement tubes are an intrinsic property of the system and  not just kinetically
trapped remainders of the preparation procedure.

\section{Simulation}
\begin{figure}
    \onefigure{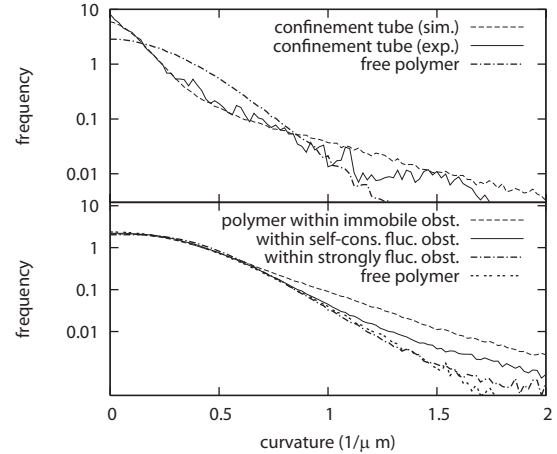}
    \caption{{\it Top:} Curvature distribution of confinement tubes of
      probe filaments in a dynamic network (experimental data and
      simulations) and distribution function of a free polymer
      (simulations). {\it Bottom:} Simulated curvature distributions
      of free filaments, filaments within a self-consistently
      fluctuating network, within a network of fixed and a network of
      strongly fluctuating obstacles.  }
    \label{curv_theory}
  \end{figure}

We resorted to Monte Carlo simulations to elucidate the underlying causes for the
observed exponential distribution of tube curvatures.
  We considered an initially equilibrated test polymer in a
  two-dimensional observation plane surrounded by randomly placed
  point-like obstacles representing the intersection points of
  neighboring polymers with that very plane.
   The tube contour was obtained by
  observing the test polymer's fluctuations in the presence of
  harmonically fluctuating point obstacles. Since these obstacles
  represent
  themselves polymers with identical properties, we tuned the
  fluctuation amplitudes of the obstacles to equal those of the test
  polymer to obtain a fully self-consistent model. Sampling over
  filaments in different network configurations and averaging over all
  statistically allowed conformations in a given tube resulted in a
  curvature distribution that also featured a pronounced exponential
  tail. As this tail shows no significant dependence on concentration
  in the experimental parameter range we compare the simulation results to the
  combined experimental data for all concentrations and find good agreement;  see Fig.~\ref{curv_theory} {\it top}.

We therefore concluded that the observed exponential distribution of
  tube curvatures is indeed characteristic for semi-dilute solutions
  of semiflexible polymers. Compared to free filaments, both smaller
  and larger curvatures are more frequent. An increased frequency of
  small curvatures is due to the coarse-graining procedure involved in
  determining the tube contour: averaging over all topologically
  allowed polymer conformations within the tube (simulations) or
  performing a time average (experiment) results in a smoother contour
  and shifts weight of the probability distribution towards small
  curvatures.
  The origin for the relatively large number of highly
  bent filaments and concomitantly the exponential tail of the
  curvature distribution is far less obvious. To avoid complications
  from coarse-graining we analyzed the distribution function of the
  confined filament itself. Even though not easily accessible to
  experiments like the tube contour, it allows a direct comparison to
  free filaments. Again, a fully self-consistent simulation of the
  polymer configuration resulted in an exponential tail in the
  curvature distribution function, cf.  Fig.~\ref{curv_theory} {\it
    bottom}, while small curvatures remained unaffected. Monitoring
  polymer segments with unusual high curvatures during the simulation
  enabled us to attribute this effect to network environments where a
  favorably bent test polymer can protrude into a large void space
  thereby realizing a higher entropy\footnote{H. Hinsch and E. Frey,
  to be published}. Upon
  immobilizing the obstacles the effect is enhanced, while for
  strongly fluctuating obstacles it decreases, cf.\
  Fig.~\ref{curv_theory} {\it bottom}. This is due to a
  smaller effective void volume and is expected to recover the free
  polymer case when strong obstacle delocalization renders the network
  comparable to a gas.  The phenomenon bears some similarity with
  ``entropic trapping'' \cite{cates88,baumgartner1987} observed for
flexible polymers
  in random environments.

\section{Discussion}
Clearly, in the absence of excluded volume, time averaging of a single test polymer
over infinite time as well as averaging over an ensemble of polymers at a given time
both have to reproduce the Gaussian distribution of free filaments. The fact that
our results do not reflect this thermodynamic equilibrium might be disturbing at
first glance, but on closer inspection turns out to be a consequence of the tube
concept and the time scales involved.

It is important to realize that for a solution of not cross-linked polymers the tube
model itself is a concept to describe polymer dynamics and not equilibrium polymer
conformations. In polymer solutions the confinement tube of a given polymer is
defined as the space explored by the polymer's fluctuations on intermediate time
scales ranging from the moment when the polymer first experiences topological
restrictions by nearest neighbors to the time scale of tube remodelling. Both points
in time can be estimated from experiment. In dynamic light scattering a cross-over
from single filament to restricted dynamics is seen at about 10 msec
\cite{semmrich:2007}. The time scale of tube remodelling can be estimated from
experiments on actin reptation by the Sackmann group \cite{dichtl:2002,keller:2003}.
From their data we deduced reptation times of 15 min for a 1 $\mu$m filament, 9 days
for a 10 $\mu$m filament and 3 years for a 50 $\mu$m filament. As reliable curvature
determination requires a polymer length of at least 8 $\mu$m, experimental
observation of terminal relaxation by reptation is entirely impossible. In essence
the same holds for a molecular dynamics simulation (even with coarse grained models)
of the process where simulation times by far exceed the possibilities of even the
most recent supercomputers due to the enormous number of different topologies.
Interestingly, for actin filaments of some micrometer length the often neglected
treadmilling motion occurring at a rate of approximately 2~$\mu$m per hour
\cite{wegner:1982,selvewegner1986} should dominate terminal relaxation. Yet, even a
relaxation time scale of 5 hours for a 10 $\mu$m filament is far beyond the time
window of light microscopic observations on such delicate samples. From these
literature data we estimate the time window during which a confinement tube is well
defined to range from 10 msec to 5 hours for not stabilized actin filaments of 10
$\mu$m length. The latter value increases to 9 days for stabilized actin networks
where the treadmilling motion of actin is abolished by phalloidin. These values
demonstrate that actin filaments in solution are trapped in their localization tubes
during almost the entire experimentally accessible time window. Consequently, all
observations on this time scale are crucially influenced by the tube's properties.

This also holds for the average performed in both experiments and simulations where
we start from an ensemble of initial configurations of test polymers embedded in the
network, and sample their configurations over a given finite time window. Hereby we
avoid the infeasible sampling of the complete phase space with one slowly reptating
test polymer, but for each initial condition we sample only over the part of phase
space that is represented by the associated localization tube. The ensemble has to
be sufficiently large to incorporate all relevant areas of the phase space. Each
member of this ensemble may now be characterized in terms of its ``topology'' given
by the topological constraints imposed by its neighbors. These topologies represent
a partitioning of phase space. If the dynamics over the finite time window would
conserve the topology, sampling again would necessarily have to give a Gaussian
distribution. Actually, however, the polymer dynamics inside their localization
tubes as well as the Markovian dynamics employed in the simulations are metrically
transitive, i.e. the partitioning is not maintained under its dynamics. In
particular, local network configurations where large void spaces are explored as
explained above by means of reptation or ``breathing'' result in axial motion of the
polymer's ends and thus allow for a modification of the topological partitioning.
Hence, entropic trapping and the concomitantly highly bent configurations of
filaments and tubes are intimately connected to transient non-equilibrium
distribution functions that are observed on time scales well below large scale
reptation sets in.

  In summary, we studied the conformations of confinement tubes in
  entangled solutions of semiflexible polymers by direct
  visualization.  Our results show that confinement tubes are very
  real objects for these systems. The dependence of the {\it average}
  tube width on actin concentration conforms to Semenov's scaling
  prediction~\cite{semenov1986}.  In experiment and simulation we
  identified coinciding exponential tails in the curvature
  distribution of the confinement tubes. This deviation is due to
  transient entropic trapping and is observed on time scales below
  large scale reptation. These findings quantify and extend the tube
  concept and demonstrate that solutions of entangled semiflexible
  polymers are not accurately  described by standard concepts of
  equilibrium thermodynamics on experimentally relevant time scales.

\acknowledgments We thank I.\ Lauter, IBN-4, for insightful discussions
of the manuscript and control measurements.
  E.F. and H.H. acknowledge support from the DFG through SFB 486, from
  the German Excellence Initiative via the NIM program and from the
  Elite Network of Bavaria through the NBT program. M.D. is grateful
  for support from the Alexander von Humboldt Foundation.


\end{document}